# PERFORMANCE COMPARISONS OF ROUTING PROTOCOLS IN MOBILE AD HOC NETWORKS

P. Manickam[1], T. Guru Baskar [2], M.Girija [3], Dr.D.Manimegalai [4]

[1,2]Department of Applied Sciences, Sethu Institute of Technology, India
gm7576@gmail.com

[3]Department of Computer Science, The American College, India

[4]Department of Information Technology, National Engineering College, India

## ABSTRACT

*Mobile Ad hoc Network (MANET) is a collection of wireless mobile nodes that dynamically form a network temporarily without any support of central administration. Moreover, Every node in MANET moves arbitrarily making the multi-hop network topology to change randomly at unpredictable times. There are several familiar routing protocols like DSDV, AODV, DSR, etc… which have been proposed for providing communication among all the nodes in the network. This paper presents a performance comparison of proactive and reactive protocols DSDV, AODV and DSR based on metrics such as throughput, packet delivery ratio and average end-to-end delay by using the NS-2 simulator.*

## 1. INTRODUCTION

A mobile ad hoc network is a collection of wireless mobile nodes that dynamically establishes the network in the absence of fixed infrastructure [1]. One of the distinctive features of MANET is, each node must be able to act as a router to find out the optimal path to forward a packet. As nodes may be mobile, entering and leaving the network, the topology of the network will change continuously. MANETs provide an emerging technology for civilian and military applications. Since the medium of the communication is wireless, only limited bandwidth is available. Another important constraint is energy due to the mobility of the nodes in nature.

One of the important research areas in MANET is establishing and maintaining the ad hoc network through the use of routing protocols. Though there are so many routing protocols available, this paper considers DSDV, AODV and DSR for performance comparisons due to its familiarity among all other protocols. These protocols are analyzed based on the important metrics such as throughput, packet delivery ratio and average end-to-end delay and is presented with the simulation results obtained by NS-2 simulator.

In particular, Section 2 presents the related works with a focus on the evaluation of the routing protocols. Section 3 briefly discusses the MANET routing protocols classification and the functionality of the three familiar routing protocols DSDV, AODV and DSR. The simulation results and performance comparison of the three above said routing protocols are discussed in Section 4. Finally, Section 5 concludes with the comparisons of the overall performance of the three protocols DSDV, AODV and DSR based on the throughput, packet delivery ratio and average end-to-end delay metrics.

## 2. RELATED WORK

A number of routing protocols have been proposed and implemented for MANETs in order to enhance the bandwidth utilization, higher throughputs, lesser overheads per packet, minimum consumption of energy and others. All these protocols have their own advantages and disadvantages under certain circumstances. The major requirements of a routing protocol was proposed by Zuraida Binti et al.[4] that includes minimum route acquisition delay, quick routing reconfiguration, loop-free routing, distributed routing approach, minimum control overhead and scalability.

MANET Routing Protocols possess two properties such as Qualitative properties (distributed operation, loop freedom, demand based routing & security) and Quantitative properties (end-to-end throughput, delay, route discovery time, memory byte requirement & network recovery time). Obviously, most of the routing protocols are qualitatively enabled. A lot of simulation studies were carried out in the paper [2] to review the quantitative properties of routing protocols.

A number of extensive simulation studies on various MANET routing protocols have been performed in terms of control overhead, memory overhead, time complexity, communication complexity, route discovery and route maintenance[16][4]. However, there is a severe lacking in implementation and operational experiences with existing MANET routing protocols. The various types of mobility models were identified and evaluated by Tracy Camp et al. [6] because the mobility of a node will also affect the overall performance of the routing protocols. A framework for the ad hoc routing protocols was proposed by Tao Lin et al. [3] using Relay Node Set which would be helpful for comparing the various routing protocols like AODV, OLSR & TBRPF [17].

The performance of the routing protocols OLSR, AODV and DSR was examined by considering the metrics of packet delivery ratio, control traffic overhead and route length by using NS-2 simulator [19][2][20][22]. The performance of the routing protocols OLSR, AODV, DSR and TORA was also evaluated with the metrics of packet delivery ratio, end-to-end delay, media access delay and throughput by also using OPNET simulator [21][23][18].

## 3. MOBILE AD HOC NETWORK ROUTING PROTOCOLS

### 3.1. Protocol Classifications

There are many ways to classify the MANET routing protocols (Figure 1), depending on how the protocols handle the packet to deliver from source to destination. But Routing protocols are broadly classified into three types such as Proactive, Reactive and Hybrid protocols [5].

#### 3.1.1. Proactive Protocols

These types of protocols are called table driven protocols in which, the route to all the nodes is maintained in routing table. Packets are transferred over the predefined route specified in the routing table. In this scheme, the packet forwarding is done faster but the routing overhead is greater because all the routes have to be defined before transferring the packets. Proactive protocols have lower latency because all the routes are maintained at all the times.
Example protocols: DSDV, OLSR (Optimized Link State Routing)

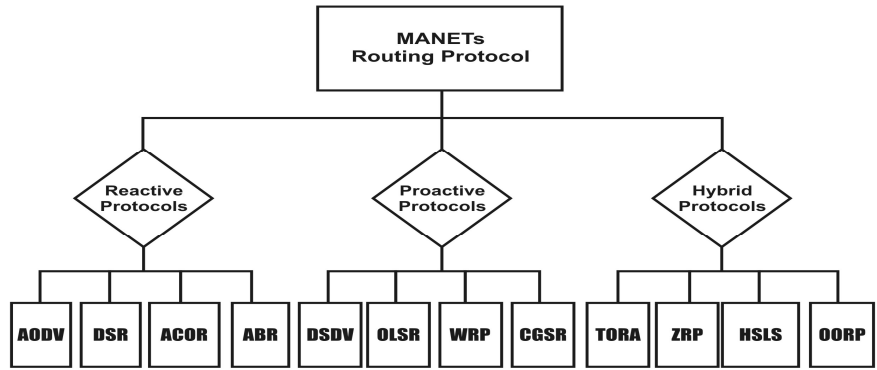

Figure 1.  MANET Routing Protocols

### 3.1.2. Reactive Protocols

These types of protocols are also called as On Demand Routing Protocols where the routes are not predefined for routing. A Source node calls for the route discovery phase to determine a new route whenever a transmission is needed. This route discovery mechanism is based on flooding algorithm which employs on the technique that a node just broadcasts the packet to all of its neighbors and intermediate nodes just forward that packet to their neighbors. This is a repetitive technique until it reaches the destination. Reactive techniques have smaller routing overheads but higher latency.
**Example Protocols:** DSR, AODV

### 3.1.3. Hybrid Protocols

Hybrid protocols are the combinations of reactive and proactive protocols and takes advantages of these two protocols and as a result, routes are found quickly in the routing zone.
**Example Protocol:** ZRP (Zone Routing Protocol)

## 3.2. Overview of Routing Protocols

In this section, a brief overview of the routing operations performed by the familiar protocols DSDV, AODV and DSR are discussed.

### 3.2.1. Destination-Sequenced Distance-Vector (DSDV) protocol

The Table-driven DSDV protocol is a modified version of the Distributed Bellman-Ford (DBF) Algorithm that was used successfully in many dynamic packet switched networks [14]. The Bellman-Ford method provided a means of calculating the shortest paths from source to destination nodes, if the metrics (distance-vectors) to each link are known. DSDV uses this idea, but overcomes DBF's tendency to create routing loops by including a parameter called destination-sequence number.

In DSDV, each node is required to transmit a sequence number, which is periodically increased by two and transmitted along with any other routing update messages to all neighboring nodes. On reception of these update messages, the neighboring nodes use the following algorithm to decide whether to ignore the update or to make the necessary changes to its routing table:

International Journal of Wireless & Mobile Networks (IJWMN) Vol. 3, No. 1, February 2011

Step 1: Receive the update message
Step 2: Update the routing table if any one of the following condition satisfies:
   i) $S_n > S_p$
  ii) $S_n = S_p$, Hop count is less
    Otherwise, ignore the update message.

Here, $S_n$ and $S_p$ are the Sequence numbers of new message and existing message respectively.

When a path becomes invalid, due to movement of nodes, the node that detected the broken link is required to inform the source, which simply erases the old path and searches for a new one for sending data. The advantages are latency for route discovery is low and loop-free path is guaranteed. The disadvantage is the huge volume of control messages.

### 3.2.2. Ad Hoc On-demand Distance Vector Routing (AODV) protocol

The Ad Hoc On-demand Distance Vector Routing (AODV) protocol is a reactive unicast routing protocol for mobile ad hoc networks [12]. As a reactive routing protocol, AODV only needs to maintain the routing information about the active paths. In AODV, the routing information is maintained in the routing tables at all the nodes. Every mobile node keeps a next-hop routing table, which contains the destinations to which it currently has a route. A routing table entry expires if it has not been used or reactivated for a pre-specified expiration time.

In AODV, when a source node wants to send packets to the destination but no route is available, it initiates a route discovery operation. In the route discovery operation, the source node broadcasts route request (RREQ) packets which includes Destination Sequence Number. When the destination or a node that has a route to the destination receives the RREQ, it checks the destination sequence numbers it currently knows and the one specified in the RREQ. To guarantee the freshness of the routing information, a route reply (RREP) packet is created and forwarded back to the source only if the destination sequence number is equal to or greater than the one specified in RREQ.

AODV uses only symmetric links and a RREP follows the reverse path of the respective RREQ. Upon receiving the RREP packet, each intermediate node along the route updates its next-hop table entries with respect to the destination node. The redundant RREP packets or RREP packets with lower destination sequence number will be dropped. The advantage of this protocol is low Connection setup delay and the disadvantage is more number of control overheads due to many route reply messages for single route request.

### 3.2.3. Dynamic Source Routing (DSR) Protocol

The Dynamic Source Routing (DSR) is a reactive unicast routing protocol that utilizes source routing algorithm [13]. In DSR, each node uses cache technology to maintain route information of all the nodes. There are two major phases in DSR such as:
- Route discovery
- Route maintenance

When a source node wants to send a packet, it first consults its route cache [7]. If the required route is available, the source node sends the packet along the path. Otherwise, the source node initiates a route discovery process by broadcasting route request packets. Receiving a route request packet, a node checks its route cache. If the node doesn't have routing information for the requested destination, it appends its own address to the route record field of the route request packet. Then, the request packet is forwarded to its neighbors.
If the route request packet reaches the destination or an intermediate node has routing information to the destination, a route reply packet is generated. When the route reply packet is generated by the destination, it comprises addresses of nodes that have been traversed by the route request packet. Otherwise, the route reply packet comprises the addresses of nodes the route request packet has traversed concatenated with the route in the intermediate node's route cache.

Whenever the data link layer detects a link disconnection, a ROUTE_ERROR packet is sent backward to the source in order to maintain the route information. After receiving the ROUTE_ERROR packet, the source node initiates another route discovery operation. Additionally, all routes containing the broken link should be removed from the route caches of the immediate nodes when the ROUTE_ERROR packet is transmitted to the source. The advantage of this protocol is reduction of route discovery control overheads with the use of route cache and the disadvantage is the increasing size of packet header with route length due to source routing.

## 4. SIMULATION RESULTS AND PERFORMANCE COMPARISONS

### 4.1. Simulation Model

Network Simulator (Version 2.29), widely known as NS2, is simply an event driven simulation tool that has proved useful in studying the dynamic nature of communication networks. Simulation of wired as well as wireless network functions and protocols (e.g., routing algorithms, TCP, UDP) can be done using NS2.

A simulation study was carried out to evaluate the performance of MANET routing protocols such as DSDV, AODV and DSR based on the metrics throughput, packet delivery ratio and average end-to-end delay with the following parameters:

| Parameter | Value |
| --- | --- |
| Radio model | TwoRay Ground |
| Protocols | DSDV,AODV,DSR |
| Traffic Source | Constant Bit Rate |
| Packet size | 512 bytes |
| Max speed | 10 m/s |
| Area | 500 x 500 |
| Number of nodes | 50, 75, 100 |
| Application | FTP |
| MAC | Mac/802_11 |
| Simulation time (Sec) | 20, 40, 60, 80 & 100 |

### 4.2. Throughput

It is the ratio of the total amount of data that reaches a receiver from a sender to the time it takes for the receiver to get the last packet. When comparing the routing throughput by each of the protocols, DSR has the high throughput. It measures of effectiveness of a routing protocol. The throughput values of DSDV, AODV and DSR Protocols for 50, 75 and 100 Nodes at Pause time 20s, 40s, 60s, 80s and 100s are noted in Table-1 and they are plotted on the different scales to best show the effects of varying throughput of the above routing protocols (Figures 2, 3 & 4).
Based on the simulation results, the throughput value of DSDV increases initially and reduces when the time increases. The throughput value of AODV slowly increases initially and maintains its value when the time increases. AODV performs well than DSDV since AODV is an on-demand protocol. The throughput value of DSR increases at lower pause time and grows as the time increases. Hence, DSR shows better performance with respect to throughput among these three protocols.

| Pause Time (Sec.) | Protocol | | | | | | | | |
|---|---|---|---|---|---|---|---|---|---|
| | DSDV | | | AODV | | | DSR | | |
| | 50 N | 75 N | 100 N | 50 N | 75 N | 100 N | 50 N | 75 N | 100 N |
| 20 | 314933 | 304192 | 1738.67 | 599851 | 692565 | 691485 | 680597 | 680597 | 680597 |
| 40 | 326862 | 315232 | 90380.9 | 547095 | 581015 | 587314 | 579319 | 575991 | 579794 |
| 60 | 230359 | 207078 | 57521.5 | 474272 | 495703 | 499404 | 492096 | 490886 | 493155 |
| 80 | 260288 | 242423 | 127322 | 439949 | 455665 | 458831 | 451614 | 450615 | 452834 |
| 100 | 276990 | 260298 | 166929 | 419988 | 432664 | 435074 | 428177 | 426776 | 429315 |

Table 1. Comparison of Throughput

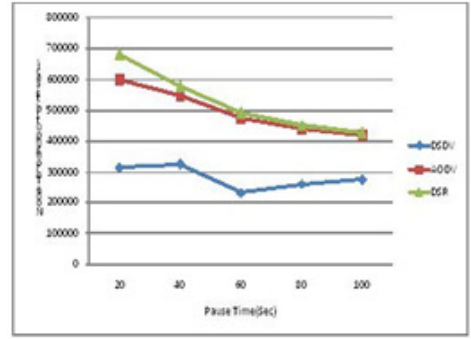

Figure 2. Comparison of Node Throughput for 50 Nodes

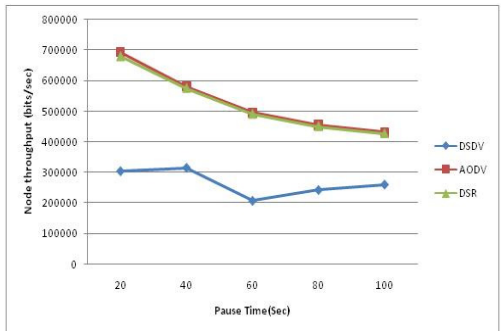

Figure 3. Comparison of Node Throughput for 75 Nodes

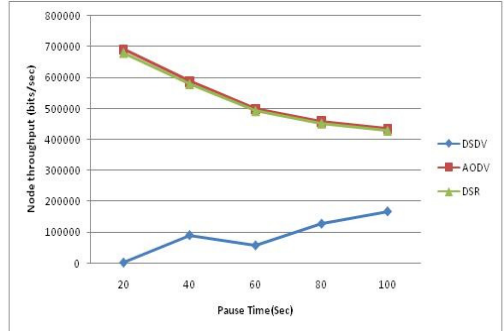

Figure 4. Comparison of Node Throughput for 100 Nodes

### 4.3. Packet delivery Ratio

Packet Delivery Ratio (PDR) is the ratio between the number of packets transmitted by a traffic source and the number of packets received by a traffic sink. It measures the loss rate as seen by transport protocols and as such, it characterizes both the correctness and efficiency of ad hoc routing protocols. A high packet delivery ratio is desired in any network.

The ratio of the Originated applications' data packets of each protocol which was able to deliver at varying time are shown in Figures 5,6 & 7 as per Table 2. As packet delivery ratio shows both the completeness and correctness of the routing protocol and also measure of efficiency the PDR value of AODV is higher than all other protocols. The PDR values of DSR and AODV are higher than that of DSDV. The PDR value of DSDV is worse in lower pause time and gradually grows in higher pause time. From the above study, in view of packet delivery ratio, reliability of AODV and DSR protocols is greater than DSDV protocol.

| Pause Time (Sec.) | Protocol | | | | | | | | |
|---|---|---|---|---|---|---|---|---|---|
| | DSDV | | | AODV | | | DSR | | |
| | 50 N | 75 N | 100 N | 50 N | 75 N | 100 N | 50 N | 75 N | 100 N |
| 20 | 97.6169 | 96.8661 | 80 | 99.0667 | 99.061 | 99.1886 | 99.1919 | 99.1909 | 99.1896 |
| 40 | 98.8569 | 98.5653 | 96.6102 | 99.1201 | 99.1093 | 99.1795 | 99.2434 | 99.2213 | 99.2031 |
| 60 | 98.4053 | 98.1191 | 96.4844 | 99.3528 | 99.3466 | 99.3854 | 99.4335 | 99.4166 | 99.404 |
| 80 | 98.8518 | 97.9306 | 97.2525 | 99.488 | 99.4843 | 99.5086 | 99.5467 | 99.5335 | 99.5233 |
| 100 | 98.4413 | 98.0971 | 97.4224 | 99.5764 | 99.5739 | 99.5307 | 99.6223 | 99.6113 | 99.6028 |

Table 2. Packet Delivery Ratio

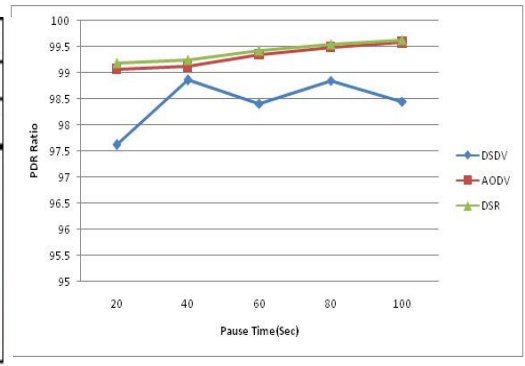

Figure 5. Comparison of PDR for 50 Nodes

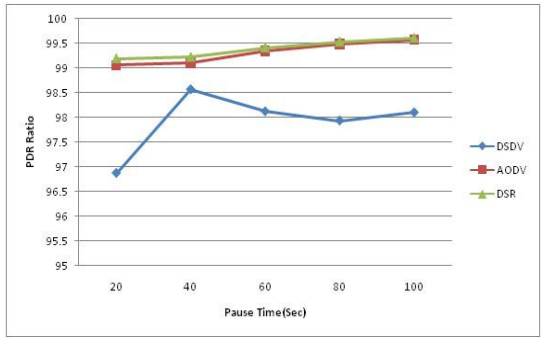 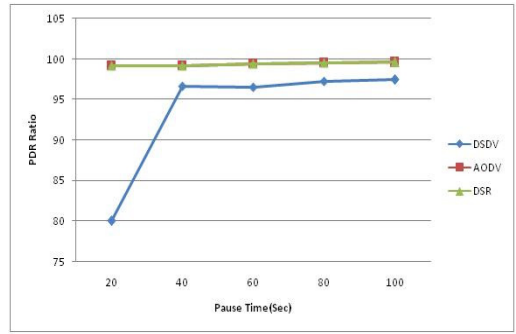

Figure 6. Comparison of PDR for 75 Nodes    Figure 7. Comparison of PDR for 100 Nodes

### 4.4. Average End-to-End delay

The packet End-to-End delay is the average time that a packet takes to traverse the network. This is the time from the generation of the packet in the sender up to its reception at the destination's application layer and it is measured in seconds. It therefore includes all the delays in the network such as buffer queues, transmission time and delays induced by routing activities and MAC control exchanges.

Various applications require different levels of packet delay. Delay sensitive applications such as voice require a low average delay in the network whereas other applications such as FTP may be tolerant to delays up to a certain level. MANETs are characterized by node mobility, packet retransmissions due to weak signal strengths between nodes, and connection tearing and making. These cause the delay in the network to increase. The End-to-End delay is therefore a measure of how well a routing protocol adapts to the various constraints in the network and represents the reliability of the routing protocol.

The Figures 8,9 &10 depict the average End-to-End delay for the DSDV, AODV and DSR protocols for the number of nodes 50, 75 & 100 respectively as per Table 3. It is clear that DSDV has the shortest End-to-End delay than AODV and DSR, because DSDV is a proactive protocol i.e. all routing informations are already stored in table. Hence, it consumes lesser time than others. On average case, DSR shows better performance than AODV but worse than DSDV. As AODV needs more time in route discovery, it produces more End-to-End delay. From the above study on End-to-End delay, DSDV has high reliability than AODV and DSR.

| Pause Time (Sec.) | Protocol | | | | | | | | |
|---|---|---|---|---|---|---|---|---|---|
| | DSDV | | | AODV | | | DSR | | |
| | 50 N | 75 N | 100 N | 50 N | 75 N | 100 N | 50 N | 75 N | 100 N |
| 20 | 0.12090 | 0.12271 | 0.32939 | 0.19027 | 0.15404 | 0.17863 | 0.09408 | 0.16907 | 0.08187 |
| 40 | 0.08895 | 0.11878 | 0.12486 | 0.17764 | 0.15607 | 0.17458 | 0.11929 | 0.16137 | 0.10740 |
| 60 | 0.09035 | 0.11878 | 0.16703 | 0.19782 | 0.17982 | 0.19330 | 0.16596 | 0.18714 | 0.13623 |
| 80 | 0.12211 | 0.14658 | 0.24473 | 0.20944 | 0.19398 | 0.20459 | 0.18486 | 0.20473 | 0.13837 |
| 100 | 0.13818 | 0.15047 | 0.23451 | 0.21645 | 0.20357 | 0.21338 | 0.20101 | 0.22017 | 0.14485 |

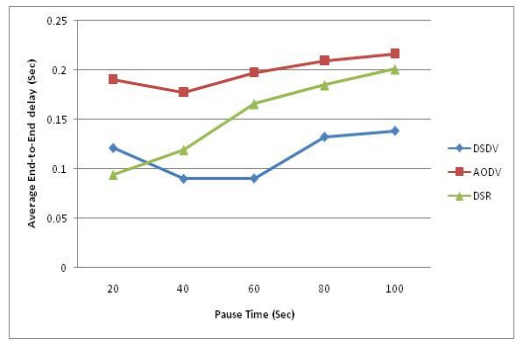

Table 3. Average End-to-End delay    Figure 8. Comparison of Average End-to-End delay for 50 Nodes

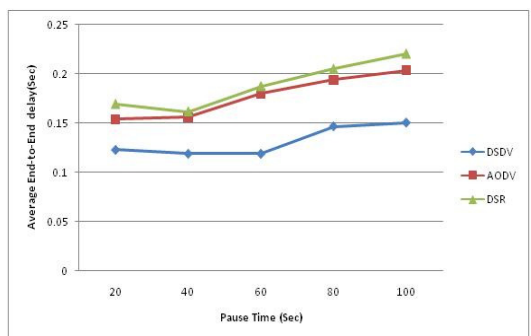
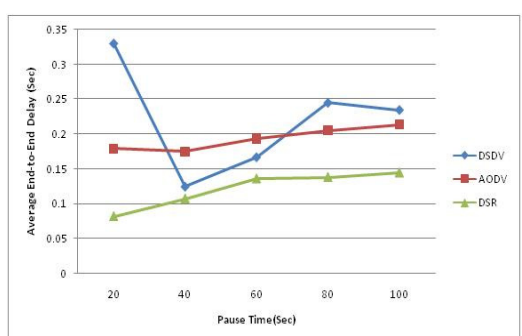

Figure 9. Comparison of Average End-to-End delay for 75 Nodes

Figure 10. Comparison of Average End-to-End delay for 100 Nodes

## 5. CONCLUSION

In this paper, the performance of the three MANET Routing protocols such as DSDV, AODV and DSR was analyzed using NS-2 Simulator. We have done comprehensive simulation results of Average End-to-End delay, throughput, and packet delivery ratio over the routing protocols DSDV, DSR and AODV by varying network size, simulation time. DSDV is a proactive routing protocol and suitable for limited number of nodes with low mobility due to the storage of routing information in the routing table at each node. Comparing DSR with DSDV and AODV protocol, byte overhead in each packet will increase whenever network topology changes since DSR protocol uses source routing and route cache. Hence, DSR is preferable for moderate traffic with moderate mobility. As AODV routing protocol needs to find route by on demand, End-to-End delay will be higher than other protocols. DSDV produces low end-to-end delay compared to other protocols. When the network load is low, AODV performs better in case of packet delivery ratio but it performs badly in terms of average End-to-End delay and throughput. Overall, DSR outperforms AODV because it has less routing overhead when nodes have high mobility considering the above said three metrics.

**Authors**

**P. Manickam** is working as an Assistant Professor in Sethu Institute of Technology, Tamil nadu, India. His current research focuses on Routing in Mobile Ad hoc Networks.

**T. GuruBaskar** is working as an Assistant Professor in Sethu Institute of Technology, Tamil nadu, India. His current research focuses on Security in Mobile Ad hoc Networks.

**M. Girija** is a Lecturer in The American College, Tamil nadu, India. Her area of specialization is Multicasting in Mobile Ad hoc Networks.

**Dr.D.Manimegalai** is Professor and Head, Department of Information Technology, National Engineering college, Tamil nadu, India. She has published more than fifteen research papers in national and International Journal and Conferences. Her area of specializations includes Image Processing, Web mining and Mobile Ad hoc Networks.